\documentclass[12pt]{article}
\usepackage{amsfonts,amssymb}

\def\C{\mathbb{C}}
\def\N{\mathbb{N}}
\def\Z{\mathbb{Z}}

\def\g{\mathfrak g}

\def\bq{ \begin{equation} }
\def\eq{ \end{equation} }
\def\ben{ \begin{eqnarray} }
\def\en{ \end{eqnarray} }

\def\frac#1#2{{#1\over #2}}

\def\on#1#2{\mathop{\vbox{\ialign{##\crcr\noalign{\kern2pt}
$\scriptstyle{#2}$\crcr\noalign{\kern2pt\nointerlineskip}
\kern-2pt$\hfil\displaystyle{#1}\hfil$\crcr}}}\limits}

\begin{document}

%%%%%%%%%%%%%%%%%%%%%%%%%%%%%%%%%%%%%%
\baselineskip=15pt
%\begin{flushright}
%Draft\\
%/04/2003
%\end{flushright}
\vspace{1cm} \centerline{{\LARGE \textbf {Integrable
pseudopotentials related to
 }}}
\vspace{0.3cm} \centerline{{\LARGE \textbf {elliptic curves
 }}}

\vskip1cm \hfill
\begin{minipage}{13.5cm}
\baselineskip=15pt {\bf A.V. Odesskii ${}^{1,2}$,  V.V. Sokolov
${}^{2}$}
\\ [2ex] {\footnotesize
${}^{1}$  Brock University  (Canada)
\\
${}^{2}$   Landau Institute for Theoretical Physics (Russia)
\\}
\vskip1cm{\bf Abstract}
%\baselineskippt

We construct integrable pseudopotentials with an arbitrary number of
fields in terms of elliptic generalization of hypergeometric functions in several variables. These
pseudopotentials yield some integrable (2+1)-dimensional
hydrodynamic type systems. An interesting class of integrable (1+1)-dimensional
hydrodynamic type systems is also generated by our pseudopotentials.

\end{minipage}

\vglue1cm \textbf{Address}: L.D. Landau Institute for Theoretical
Physics of Russian Academy of Sciences, Kosygina 2, 119334,
Moscow, Russia

\textbf{E-mail}: aodesski@brocku.ca, sokolov@itp.ac.ru

\newpage \tableofcontents
\newpage

\section{Introduction}

In \cite{odsok1} a wide class of 3-dimensional integrable PDEs of the form
\begin{equation}   \label{genern}
\sum_{j=1}^m a_{ij}({\bf u})\,u_{j,t_{1}}+\sum_{j=1}^m
b_{ij}({\bf u})\,u_{j, t_{2}}+ \sum_{j=1}^m c_{ij}({\bf
u})\,u_{j, t_{3}}=0, \qquad i=1,...,l,
\end{equation}
where ${\bf u}=(u_{1},\dots, u_{m})$ was constructed. The coefficients
of these PDEs were written in terms of generalized hypergeometric functions \cite{gel}.
By the integrability of (\ref{genern}) we mean the existence of a pseudopotential
representation\footnote{This means that (\ref{genern}) is equivalent to the compatibility conditions for (\ref{pseudo}).}
\begin{equation} \label{pseudo}
\psi_{t_{2}}=A(p,{\bf u}), \qquad  \psi_{t_{3}}=B(p,{\bf
u}), \qquad \mbox{where}\quad p = \psi_{t_{1}}.
\end{equation}
Such a pseudopotential representation is a dispersionless version \cite{Zakharov, kr3} of the zero curvature
representation, which is a basic notion in the integrability theory
of solitonic equations (see \cite{zahshab}). One of the interesting and attractive  features of the theory of
integrable systems (\ref{genern}) is that the dependence of the
pseudopotentials  on $p$ can be much more
complicated then in the solitonic case. In \cite{kr4,
dub} some important examples of pseudopotentials $A,B$
related to the Whitham averaging procedure for integrable dispersion
PDEs and to the Frobenious manifolds  were found. These examples are related to the universal algebraic curve of
genus $g$ with $M$ punctures for arbitrary $g,~M$. More precisely, the point
$\left(\frac{A_{ppp}}{A_{pp}^2},A_p\right)$ runs over a curve of genus $g$ while $p$ runs over $\C$ and ${\bf u}$
are some coordinates on the moduli space ${\cal M}_{g,M}$ of curves of genus $g$ with $M$ punctures.

The pseudopotentials from \cite{odsok1} (see also \cite{odes}) were written in the following
parametric form:
$$A=F_1(\xi, {\bf u}),\qquad p=F_2(\xi, {\bf u}),$$
where the $\xi$-dependence of the functions $F_i$ is defined by the
ODE
\begin{equation}\label{ODE}
\label{par}F_{i,\xi}=\phi_i(\xi,{\bf u})\cdot\xi^{-s_1}(\xi-1)^{-s_2}(\xi-u_1)^{-s_3}...(\xi-u_m)^{-s_{m+2}}.
\end{equation}
Here $s_1,...,s_{m+2}$ are arbitrary constants and $\phi_i$ are
polynomials  in $\xi$ of degree $m-k$.  These pseudopotentials are related to rational
algebraic curves. If $s_1=...=s_{m+2}=0$ and $k=0,$ then they coincide with pseudopotentials from
\cite{kr4} related to ${\cal M}_{0,m+3}$.

In this paper we construct integrable systems (\ref{genern}) and pseudopotentials related to elliptic curve.
For these systems  ${\bf u}=(u_{1},\dots, u_{n},\tau)$, where $\tau$ is the parameter of the elliptic
curve. Note that $\tau$ is also an unknown function in our systems (\ref{genern}). The coefficients of the systems
are expressed in terms of some elliptic generalization of hypergeometric
functions in several variables. These elliptic  hypergeometric functions can be defined as solutions of the following compatible linear
overdetermined system of PDEs:
\begin{equation}\begin{array}{c}
\label{elldarbu}
\displaystyle g_{u_{\alpha}u_{\beta}}=s_{\beta}\Big(\rho(u_{\beta}-u_{\alpha})
+\rho(u_{\alpha}+\eta)-\rho(u_{\beta})
-\rho(\eta)\Big)g_{u_{\alpha}}+\\[3mm]
\displaystyle s_{\alpha}\Big(\rho(u_{\alpha}-u_{\beta})
+\rho(u_{\beta}+\eta)-\rho(u_{\alpha})
-\rho(\eta)\Big)g_{u_{\beta}},
\\[7mm]
\displaystyle g_{u_{\alpha}u_{\alpha}}=s_{\alpha}\sum_{\beta\ne\alpha}\Big(\rho(u_{\alpha})+
\rho(\eta)-\rho(u_{\alpha}-u_{\beta})-\rho(u_{\beta}+\eta)\Big)g_{u_{\beta}}+
\\[5mm]
\displaystyle \Big(\sum_{\beta\ne\alpha}s_{\beta}\rho(u_{\alpha}-u_{\beta})
+(s_{\alpha}+1)\rho(u_{\alpha}+\eta)+
s_{\alpha}\rho(-\eta)+(s_0-s_{\alpha}-1)\rho(u_{\alpha})+2\pi ir \Big)g_{u_{\alpha}}-
\\[3mm]
\displaystyle s_0s_{\alpha}(\rho^{\prime}(u_{\alpha})-\rho^{\prime}(\eta))g,
\\[7mm]
\displaystyle g_{\tau}=\frac{1}{2\pi i}\sum_{\beta}\Big(\rho(u_{\beta}+\eta)-
\rho(\eta)\Big)g_{u_{\beta}}-\frac{s_0}{2\pi
i}\rho^{\prime}(\eta) g
\end{array}
\end{equation}
for a single function
$g(u_{1},\dots,u_{n},\tau).$
Here and in the sequel $\eta=s_1u_1+...+s_nu_n+r\tau+\eta_0,~s_0=-s_1-...-s_n,$ where $s_1,...,s_n,r,\eta_0$ are arbitrary
constants, and
\begin{equation}\label{teta}
\theta(z)=\sum_{\alpha\in\Z}(-1)^{\alpha}e^{2\pi i(\alpha z+\frac{\alpha(\alpha-1)}{2}\tau)}, \qquad
\rho(z)=\frac{\theta^{\prime}(z)}{\theta(z)}.
\end{equation}
In the above formulas and in the sequel we omit the second argument $\tau$ of the functions $\theta$, $\rho$ and use the notation
$$\rho^{\prime}(z)=\frac{\partial\rho(z)}{\partial z},\qquad \rho_{\tau}(z)=\frac{\partial\rho(z)}{\partial \tau},\qquad \theta^{\prime}(z)=\frac{\partial\theta(z)}{\partial z},\qquad \theta_{\tau}(z)=\frac{\partial\theta(z)}{\partial \tau}.$$
It turns out that the dimension of the space of solutions for
(\ref{elldarbu}) equals $n+1$.

The paper is organized as follows.

In Section 2 we describe some properties of elliptic hypergeometric functions needed for our
purposes. In particular, we present an integral representation similar to
the representation for the generalized hypergeometric function
(see, for example \cite{odsok1}).

In Section 3  for any $n$ we construct pseudopotentials (\ref{pseudo}) with $k=0$ related to the elliptic hypergeometric
functions. The pseudopotential $A_n(p,u_1,...,u_n,\tau)$ is defined in a parametric form by
\begin{equation}\label{psdef0}
A_n=P_n(g_1,\xi), \qquad p=P_n(g_0,\xi),
\end{equation}
where $g_1,~g_0$ be linearly independent solutions of (\ref{elldarbu}),
\begin{equation}\label{pp}P_n(g,\xi)=\int_0^{\xi}S_n(g,\xi)e^{2\pi ir(\tau-\xi)}
\frac{\theta^{\prime}(0)^{-s_1-...-s_n}\theta(u_1)^{s_1}...\theta(u_n)^{s_n}}
{\theta(\xi)^{-s_1-...-s_n}\theta(\xi-u_1)^{s_1}...\theta(\xi-u_n)^{s_n}}
d\xi,
\end{equation}
and
\begin{equation}\label{ell0}
S_n(g, \xi)=\sum_{1\leq \alpha\leq
n}\frac{\theta(u_{\alpha})\theta(\xi-u_{\alpha}-\eta)}{\theta(u_{\alpha}+\eta)
\theta(\xi-u_{\alpha})}g_{u_{\alpha}}-(s_1+...+s_n)\frac{\theta^{\prime}(0)\theta(\xi-\eta)}{\theta(\eta)\theta(\xi)}g.
\end{equation}
We call them {\it elliptic pseudopotential of defect 0}. Such pseudopotentials
define integrable systems of the form (\ref{genern}) with $m=l=n+1.$ In the
case $s_1=...=s_n=r=0,~ \eta_{0} \rightarrow 0 $
our pseudopotentials   coincide with elliptic
pseudopotentials constructed in \cite{kr4}.

In Section 4 for $k<n$ we construct  {\it
pseudopotentials of defect $k$}. These pseudopotentials
define  systems   (\ref{genern}) with $m=n+1,\,l=n+k+1.$

A special class of solutions  for integrable systems  (\ref{genern}) depending on several arbitrary functions of one variable can be
constructed by the method of hydrodynamic reductions \cite{Gibt,ferhus1}.
The hydrodynamic reductions are defined by pairs of integrable compatible
(1+1)-dimensional hydrodynamic type systems of the form
\begin{equation}\label{gidra}
r_{t}^{i}=v^{i}(r^1,...,r^N)r_{x}^{i},\qquad i=1,2,...,N.
\end{equation}
These integrable systems also are of interest themselves. A general theory of such type integrable
systems was developed in \cite{Dubrovin2,tsar}.

Section 5 is devoted to hydrodynamic reductions of systems (\ref{genern}) constructed in Sections 3,4. In the case $k=0$
the corresponding systems
(\ref{gidra}) are defined by
\begin{equation}\label{gidragen1}
r^i_t=\frac{S_{n}(g_1({\bf u}),\xi_i)}{S_{n}(g_2({\bf u}),\xi_i)}r^i_x,
\end{equation}
where $g_{1},g_{2}$ are linearly independent solutions of
(\ref{elldarbu}). For the pseudopotentials of defect $k>0$ the corresponding formula is similar.
The functions
$\tau(r^1,...,r^N)$,$~\xi_i(r^1,...,r^N),$ $u_i(r^1,...,r^N)$
are defined by the following
universal overdetermined compatible system of PDEs of the
Gibbons-Tsarev type \cite{Gibt, pav2}:
\begin{equation}\label{gibtsar1}
\partial_{\alpha}\xi_{\beta}=\frac{1}{2\pi i}\Big(\rho(\xi_{\alpha}-\xi_{\beta})-\rho(\xi_{\alpha})\Big)\partial_{\alpha}
\tau, \qquad  \partial_{\alpha}=\frac{\partial}{\partial r_{\alpha}},
\end{equation}
\begin{equation}\label{gibtsar11}
\partial_{\alpha} \partial_{\beta} \tau=-\frac{1}{\pi i}\rho^{\prime}(\xi_{\alpha}-\xi_{\beta})\partial_{\alpha} \tau \partial_{\beta} \tau,
\end{equation}
and
\begin{equation}\label{u1}
\partial_{\alpha}u_{\beta}=\frac{1}{2\pi i}\Big(\rho(\xi_{\alpha}-u_{\beta})-\rho(\xi_{\alpha})\Big)\partial_{\alpha}
\tau,\qquad \alpha=1,...,N,~~~\beta=1,...,n.
\end{equation}
Recall that here $\tau$ is the second argument of the function $\rho$. It would be interesting to compare formulas
(\ref{gibtsar1}), (\ref{gibtsar11}) with formulas (3.23)-(3.26) from
\cite{vasil}.

It is easy to verify that the system (\ref{gibtsar1})-(\ref{u1}) is
consistent. Therefore our (1+1)-dimensional systems (\ref{gidragen1})
admit a local parameterization by $2 N$ arbitrary functions of one variable.

For some very special values of parameters $s_{\alpha}$ in (\ref{elldarbu}) our systems (\ref{gidragen1}) are related to the Whitham
hierarchies \cite{kr4}, to the Frobenious manifolds \cite{dub,kr},
and to the associativity equation \cite{dub,kr}.

\section{Elliptic hypergeometric functions}

Define a function $\theta$ in two variables $z, ~\tau$ by (\ref{teta}).
We assume that ${\rm Im} \tau>0$. The function $\theta$ is called theta-function of order one in one variable.
Recall the following useful formulas:
$$\theta(z+1)=\theta(z)\qquad \theta(z+\tau)=-e^{-2\pi i z}\theta(z),$$
$$\theta(-z)=-e^{-2\pi i z}\theta(z),\qquad \theta_{\tau}(z)=\frac{1}{4\pi i}\theta^{\prime\prime}(z)-\frac{1}{2}\theta^{\prime}(z).$$

The following statements can be verified straightforwardly.

{\bf Proposition 1.} The system of linear equations (\ref{elldarbu}) is compatible for any constants $s_1,\dots, s_n,r,\eta_0.$ The
dimension of  the linear space  ${\cal H}$ of
solutions for system (\ref{elldarbu}) is equal to
$n+1$. $\blacksquare$

{\bf Remark 1.} Some coefficients of (\ref{elldarbu}) can be written in a
factorized form using the following identity
$$\rho(u_{\beta}-u_{\alpha})+\rho(u_{\alpha}+\eta)-\rho(u_{\beta})-\rho(\eta)
=-\frac{\theta^{\prime}(0)\theta(u_{\alpha}-u_{\beta}+\eta)\theta(u_{\alpha})\theta(u_{\beta}+\eta)}{\theta(u_{\alpha}
-u_{\beta})\theta(u_{\alpha}+\eta)\theta(u_{\beta})\theta(\eta)}.$$

We call elements of  ${\cal H}$ {\it elliptic hypergeometric functions.}

{\bf Proposition 2.} Define a function $F(u_1,...,u_n,\tau)$ by
the following integral representation
$$
F(u_1,...,u_n,\tau)=  \int_0^1\frac{\theta(u_1-t)^{s_1}...\theta(u_n-t)^{s_n}\theta^{\prime}(0)^{s_1+...+s_n+1}\theta(t+\eta)}{\theta(u_1)^{s_1}...\theta(u_n)^{s_n}\theta(t)^{s_1+...+s_n+1}\theta(\eta)}e^{2\pi i(s_1+...+s_n+r)t}
dt.
$$
Then the function $F$ satisfies system (\ref{elldarbu}).

{\bf Proposition 3.} Let ${\cal H}={\cal H}_{s_1,...,s_{n},r,\eta_0}$ and
$\widetilde{{\cal H}}={\cal H}_{s_1,...,s_n,0,r,\eta_0}$.
Then $\widetilde{{\cal H}}$ is spanned by ${\cal H}$ and by the
function
\begin{equation}\label{gen}
Z(u_1,...,u_n,u_{n+1},\tau)=\int_0^{u_{n+1}}\frac{\theta(u_1-t)^{s_1}...\theta(u_n-t)^{s_n}\theta^{\prime}(0)^{s_1+...+s_n+1}\theta(t+\eta)}{\theta(u_1)^{s_1}...\theta(u_n)^{s_n}\theta(t)^{s_1+...+s_n+1}\theta(\eta)}e^{2\pi i(s_1+...+s_n+r)t}dt.
\end{equation}
Moreover, the space ${\cal H}_{s_1,...,s_n,0,...,0,r,\eta_0}$
($m$ zeros) is spanned by ${\cal H}$ and by $Z(u_1,...,u_n,u_{n+1},\tau)$,
$Z(u_1,...,u_n,u_{n+2},\tau), ..., Z(u_1,...,u_n,u_{n+m},\tau)$.
$\blacksquare$

In the simplest case
$n=1$   system (\ref{elldarbu}) for a function $g(u_{1}, \tau)$ has the following form
$$g_{u_{1}u_{1}}=
 \Big((s_{1}+1)\frac{\theta^{\prime}(u_{1}+\eta)}{\theta(u_{1}+\eta)}+
s_{1}\frac{\theta^{\prime}(-\eta)}{\theta(-\eta)}-(2 s_{1}+1)\frac{\theta^{\prime}
(u_{1})}{\theta(u_{1})}+2\pi ir\Big) g_{u_{1}}-$$
$$\frac{s_{1}^{2}\theta^{\prime}(0)^2\theta(u_{1}-\eta)\theta(u_{1}+\eta)}
{\theta(u_{1})^2\theta(-\eta)\theta(\eta)}g,$$
$$g_{\tau}=\frac{1}{2\pi i}\Big(\frac{\theta^{\prime}(u_{1}+\eta)}{\theta(u_{1}+\eta)}-
\frac{\theta^{\prime}(\eta)}{\theta(\eta)}\Big)g_{u_{1}}+\frac{s_1}{2\pi
i}(\ln\theta(\eta))^{\prime\prime}g.
$$
If $s_{1}=0$, then the functions $g=1$ and
$$g=Z(u_1,\tau)=
\int_0^{u_1}\frac{\theta^{\prime}(0)\theta(t+\eta)}{\theta(t)\theta(\eta)}e^{2\pi irt}dt$$
span the space of solutions.
Moreover, Proposition 3
implies that the functions $1,$ $Z(u_1,\tau)$, ..., $Z(u_n,\tau)$ span the space
of solutions of (\ref{elldarbu}) in the case $s_{1}=\cdots=s_{n}=0$.

\section{Elliptic pseudopotentials of defect 0}

For any elliptic hypergeometric function $g\in\cal H$ we put
\begin{equation}\label{ell0}
S_n(g, \xi)=\sum_{1\leq \alpha\leq
n}\frac{\theta(u_{\alpha})\theta(\xi-u_{\alpha}-\eta)}{\theta(u_{\alpha}+\eta)\theta(\xi-u_{\alpha})}g_{u_{\alpha}}-(s_1+...+s_n)\frac{\theta^{\prime}(0)\theta(\xi-\eta)}{\theta(\eta)\theta(\xi)}g.
\end{equation}
Define $P_n(g,\xi)$ by the formula
(\ref{pp}) if ${\rm Re}\, (s_1+...+s_n)>1$ and
as the analytic continuation of (\ref{pp}) otherwise.

{\bf Proposition 4.} The following relations hold
\begin{equation}
\label{pr1}
(P_n(g,\xi))_{u_{\alpha}}=-\frac{g_{u_{\alpha}}\theta(u_{\alpha})\theta(\xi-u_{\alpha}-\eta)}
{\theta(u_{\alpha}+\eta)\theta(\xi-u_{\alpha})}e^{2\pi ir(\tau-\xi)}\frac{\theta^{\prime}(0)^{-s_1-...-s_n}
\theta(u_1)^{s_1}...\theta(u_n)^{s_n}}{\theta(\xi)^{-s_1-...-s_n}\theta(\xi-u_1)^{s_1}...\theta(\xi-u_n)^{s_n}},
\end{equation}
\begin{equation}
\label{pr2}
\begin{array}{c}
\displaystyle (P_n(g,\xi))_{\tau}=\Big(\frac{1}{2\pi i}(\sum_{1\leq \alpha\leq
n}\frac{\theta(u_{\alpha})\theta^{\prime}(\xi-u_{\alpha}-\eta)}
{\theta(u_{\alpha}+\eta)\theta(\xi-u_{\alpha})}g_{u_{\alpha}}-(s_1+...+s_n)
\frac{\theta^{\prime}(0)\theta^{\prime}(\xi-\eta)}{\theta(\eta)\theta(\xi)}g)-\\[6mm]
\displaystyle \frac{\theta^{\prime}(-\eta)}{2\pi i\theta(-\eta)}(\sum_{1\leq \alpha\leq
n}\frac{\theta(u_{\alpha})\theta(\xi-u_{\alpha}-\eta)}{\theta(u_{\alpha}+\eta)\theta(\xi-u_{\alpha})}g_{u_{\alpha}}
-(s_1+...+s_n)\frac{\theta^{\prime}(0)\theta(\xi-\eta)}{\theta(\eta)\theta(\xi)}g)\Big)
\times\\[6mm]
\displaystyle e^{2\pi ir(\tau-\xi)}\frac{\theta^{\prime}(0)^{-s_1-...-s_n}\theta(u_1)^{s_1}...
\theta(u_n)^{s_n}}{\theta(\xi)^{-s_1-...-s_n}\theta(\xi-u_1)^{s_1}...\theta(\xi-u_n)^{s_n}}.
\end{array}
\end{equation}

{\bf Proof.} Taking the derivatives of (\ref{pr1}), (\ref{pr2}) with respect to $\xi,$ one
arrives at theta-functional identities, which can be proved straightforwardly.
Moreover, the values of the left and the right hand sides of (\ref{pr1}) and (\ref{pr2}) are equal to zero  at
$\xi=0$.
$\blacksquare$

Let $g_1,~g_0$ be linearly independent elements of ${\cal H}$. A
pseudopotential $A_n(p,u_1,...,u_n,\tau)$ defined in a parametric form by (\ref{psdef0})
is called {\it elliptic pseudopotential of defect 0}.
Relations (\ref{psdef0}) mean that to find $A_{n}(p,u_1,...,u_n,\tau),$ one has to express $\xi$ from the second equation and
substitute the result into the first equation.

Let $g_0,g_1,...,g_n\in {\cal H}$ be a basis in ${\cal H}.$ Define
pseudopotentials $B_{\alpha}(p,u_1,...,u_n,\tau)$  of defect 0, where
$\alpha=1,...,n,$ by
\begin{equation}
\label{B}B_{\alpha}=P_n(g_{\alpha},\xi), \qquad p=P_n(g_0,\xi), \qquad \alpha=1,...,n.
\end{equation}

Suppose that $u_1,...,u_n,\tau$ are functions
of $t_0=x,~t_1,...,t_n$.

{\bf Theorem 1.}
The compatibility conditions
$\psi_{t_{\alpha}t_{\beta}}=\psi_{t_{\beta}t_{\alpha}}$ for the
system
\begin{equation}
\label{pseudopar}\psi_{t_{\alpha}}=B_{\alpha}(\psi_x,u_1,...,u_n,\tau),
\qquad \alpha=1,...,n,
\end{equation}
are equivalent to the following system of PDEs for $u_1,...,u_n,\tau$:
$$\sum_{1\leq\beta\leq n}(g_qg_{r,u_{\beta}}-g_rg_{q,u_{\beta}})(u_{\beta,t_s}+\frac{1}{2\pi i}(\frac{\theta^{\prime}(u_{\beta}+\eta)}{\theta(u_{\beta}+\eta)}-\frac{\theta^{\prime}(\eta)}{\theta(\eta)})\tau_{t_s})+$$
\begin{equation}\label{sys1}\sum_{1\leq\beta\leq n}(g_rg_{s,u_{\beta}}-g_sg_{r,u_{\beta}})(u_{\beta,t_q}+\frac{1}{2\pi i}(\frac{\theta^{\prime}(u_{\beta}+\eta)}{\theta(u_{\beta}+\eta)}-\frac{\theta^{\prime}(\eta)}{\theta(\eta)})\tau_{t_q})+
\end{equation}
$$\sum_{1\leq\beta\leq n}(g_sg_{q,u_{\beta}}-g_qg_{s,u_{\beta}})(u_{\beta,t_r}+\frac{1}{2\pi i}(\frac{\theta^{\prime}(u_{\beta}+\eta)}{\theta(u_{\beta}+\eta)}-\frac{\theta^{\prime}(\eta)}{\theta(\eta)})\tau_{t_r})=0,$$
$$\sum_{1\leq \beta\leq n,\beta\ne
\alpha}\frac{\theta(u_{\beta})\theta(u_{\alpha}-u_{\beta}-\eta)}{\theta(u_{\beta}+\eta)\theta(u_{\alpha}-u_{\beta})}(g_{r,u_{\alpha}}g_{q,u_{\beta}}-g_{q,u_{\alpha}}g_{r,u_{\beta}})(u_{\alpha,t_s}-u_{\beta,t_s}+\frac{1}{2\pi i}(\frac{\theta^{\prime}(u_{\alpha}-u_{\beta}-\eta)}{\theta(u_{\alpha}-u_{\beta}-\eta)}-\frac{\theta^{\prime}(-\eta)}{\theta(-\eta)})\tau_{t_s})+$$
$$
\sum_{1\leq \beta\leq n,\beta\ne
\alpha}\frac{\theta(u_{\beta})\theta(u_{\alpha}-u_{\beta}-\eta)}{\theta(u_{\beta}+\eta)\theta(u_{\alpha}-u_{\beta})}(g_{s,u_{\alpha}}g_{r,u_{\beta}}-g_{r,u_{\alpha}}g_{s,u_{\beta}})(u_{\alpha,t_q}-u_{\beta,t_q}+\frac{1}{2\pi i}(\frac{\theta^{\prime}(u_{\alpha}-u_{\beta}-\eta)}{\theta(u_{\alpha}-u_{\beta}-\eta)}-\frac{\theta^{\prime}(-\eta)}{\theta(-\eta)})\tau_{t_q})+
$$
$$\sum_{1\leq \beta\leq n,\beta\ne
\alpha}\frac{\theta(u_{\beta})\theta(u_{\alpha}-u_{\beta}-\eta)}{\theta(u_{\beta}+\eta)\theta(u_{\alpha}-u_{\beta})}(g_{q,u_{\alpha}}g_{s,u_{\beta}}-g_{s,u_{\alpha}}g_{q,u_{\beta}})(u_{\alpha,t_r}-u_{\beta,t_r}+\frac{1}{2\pi i}(\frac{\theta^{\prime}(u_{\alpha}-u_{\beta}-\eta)}{\theta(u_{\alpha}-u_{\beta}-\eta)}-\frac{\theta^{\prime}(-\eta)}{\theta(-\eta)})\tau_{t_r})-$$
$$(s_1+...+s_n)\frac{\theta^{\prime}(0)\theta(u_{\alpha}-\eta)}{\theta(\eta)\theta(u_{\alpha})}(g_qg_{r,u_{\alpha}}-g_rg_{q,u_{\alpha}})(u_{\alpha,t_s}+\frac{1}{2\pi i}(\frac{\theta^{\prime}(u_{\alpha}-\eta)}{\theta(u_{\alpha}-\eta)}-\frac{\theta^{\prime}(-\eta)}{\theta(-\eta)})\tau_{t_s})-$$
\begin{equation}\label{sys2}(s_1+...+s_n)\frac{\theta^{\prime}(0)\theta(u_{\alpha}-\eta)}{\theta(\eta)\theta(u_{\alpha})}(g_rg_{s,u_{\alpha}}-g_sg_{r,u_{\alpha}})(u_{\alpha,t_q}+\frac{1}{2\pi i}(\frac{\theta^{\prime}(u_{\alpha}-\eta)}{\theta(u_{\alpha}-\eta)}-\frac{\theta^{\prime}(-\eta)}{\theta(-\eta)})\tau_{t_q})-
\end{equation}
$$(s_1+...+s_n)\frac{\theta^{\prime}(0)\theta(u_{\alpha}-\eta)}{\theta(\eta)\theta(u_{\alpha})}(g_sg_{q,u_{\alpha}}-g_qg_{s,u_{\alpha}})(u_{\alpha,t_r}+\frac{1}{2\pi i}(\frac{\theta^{\prime}(u_{\alpha}-\eta)}{\theta(u_{\alpha}-\eta)}-\frac{\theta^{\prime}(-\eta)}{\theta(-\eta)})\tau_{t_r})=0,$$ where
$\alpha=1,...,n$. Here $q, r, s$ run from 0
to $n$.

{\bf Proof.} Taking into account (\ref{B}), we find that the compatibility conditions for (\ref{pseudopar}) are equivalent to
$$\sum_{\alpha=1}^n\Big(((P_n(g_q,\xi))_{\xi}(P_n(g_r,\xi))_{u_{\alpha}}-(P_n(g_r,\xi))_{\xi}(P_n(g_q,\xi))_{u_{\alpha}})u_{\alpha,t_{s}}+$$
$$((P_n(g_r,\xi))_{\xi}(P_n(g_s,\xi))_{u_{\alpha}}-(P_n(g_s,\xi))_{\xi}(P_n(g_r,\xi))_{u_{\alpha}})u_{\alpha,t_{q}}+$$
\begin{equation}
\label{whit}((P_n(g_s,\xi))_{\xi}(P_n(g_q,\xi))_{u_{\alpha}}-(P_n(g_q,\xi))_{\xi}(P_n(g_s,\xi))_{u_{\alpha}})u_{\alpha,t_{r}}\Big)+
\end{equation}
$$((P_n(g_q,\xi))_{\xi}(P_n(g_r,\xi))_{\tau}-(P_n(g_r,\xi))_{\xi}(P_n(g_q,\xi))_{\tau})\tau_{t_{s}}+$$
$$((P_n(g_r,\xi))_{\xi}(P_n(g_s,\xi))_{\tau}-(P_n(g_s,\xi))_{\xi}(P_n(g_r,\xi))_{\tau})\tau_{t_{q}}+$$
$$((P_n(g_s,\xi))_{\xi}(P_n(g_q,\xi))_{\tau}-(P_n(g_q,\xi))_{\xi}(P_n(g_s,\xi))_{\tau})\tau_{t_{r}}=0.$$
Using (\ref{pp}), (\ref{pr2}), we rewrite (\ref{whit}) as follows:
$$\sum_{1\leq \beta\leq n}\frac{\theta(u_{\beta})\theta(\xi-u_{\beta}-\eta)}
{\theta(u_{\beta}+\eta)\theta(\xi-u_{\beta})}(S_n(g_q,\xi)g_{r,u_{\beta}}-S_n(g_r,\xi)g_{q,u_{\beta}})u_{\beta,t_s}-$$
\begin{equation}
\label{whit1}\frac{1}{2\pi i}\sum_{1\leq \beta\leq n}\frac{\theta(u_{\beta})\theta^{\prime}(\xi-u_{\beta}-\eta)}
{\theta(u_{\beta}+\eta)\theta(\xi-u_{\beta})}(S_n(g_q,\xi)g_{r,u_{\beta}}-S_n(g_r,\xi)g_{q,u_{\beta}})\tau_{t_s}+
\end{equation}
$$\frac{1}{2\pi i}(s_1+...+s_n)\frac{\theta^{\prime}(0)\theta^{\prime}(\xi-\eta)}{\theta(\eta)\theta(\xi)}
(S_n(g_q,\xi)g_r-S_n(g_r,\xi)g_q)\tau_{t_s}+(q,r,s)=0,$$
where $(q,r,s)$ means the cyclic permutation of $q,r,s$. Denote the left hand side of (\ref{whit1}) by $\Lambda(\xi)$. One can check that
$$\Lambda(\xi+1)=\Lambda(\xi),\qquad \Lambda(\xi+\tau)=e^{4\pi i\eta}\Lambda(\xi),$$
and the only singularities of $\Lambda(\xi)$ are poles of order one at the points $\xi=0,u_1,...,u_n$ modulo $1,~\tau$.
This implies that $\Lambda(\xi)=0$ iff the residues at these points are equal to zero.
Calculating the residue at $\xi=0$, we get (\ref{sys1}). The calculation of the residue at $\xi=u_{\alpha}$ leads to (\ref{sys2}).
$\blacksquare$

{\bf Remark 2.} Given $t_{1},t_{2},t_{3},$ Theorem 1 yields a
3-dimensional system of the form (\ref{genern}) with $l=m=n+1$
possessing a pseudopotential representation.

{\bf Remark 3.} Consider the case $s_1=...=s_n=0$. We have
$$g=c_0+c_1Z(u_1,\tau)+...+c_nZ(u_n,\tau),$$
where $c_0,...,c_n$ are constants. Therefore,
$$S_n(g,\xi)=\sum_{1\leq\alpha\leq n}c_{\alpha}e^{2\pi iru_{\alpha}}\frac{\theta^{\prime}
(0)\theta(\xi-u_{\alpha}-\eta_0)}{\theta(\eta_0)\theta(\xi-u_{\alpha})}.$$
If we assume $r=0$ and $c_1+...+c_n=0$, then in the limit $\eta_{0} \rightarrow 0$ we obtain
$$S_n(g,\xi)=\sum_{1\leq\alpha\leq n}c_{\alpha}\rho(\xi-u_{\alpha}).$$
 A system of PDEs
equivalent to compatibility conditions for equations of the form
(\ref{whit}), was called  in \cite{kr4} a {\it Whitham hierarchy}.
In this paper I.M. Krichever constructed some Whitham
hierarchies related to algebraic curves of arbitrary genus $g$. The
hierarchy corresponding to $g=1$ is equivalent to one described by
Theorem 1 if $r=s_{1}=\dots=s_n=0,$  $c_1+..+c_n=0$, and $\eta_0\to0$ as described above.

\section{Elliptic pseudopotentials of defect $k>0$}

In this section we construct {\it elliptic pseudopotentials of defect} $k.$ Fix $k$ linearly independent
elliptic hypergeometric functions $h_1,...,h_k\in {\cal H}$. For
any $g\in {\cal H}$ define $P_{n,k}(g,\xi)$ by the formula
\begin{equation}\label{polgen}
P_{n,k}(g,\xi)=\frac{1}{\Delta}\det\left(\begin{array}{cccc}P_n(g,\xi)&P_n(h_1,\xi)&...&P_n(h_k,\xi)
\\g_{u_{n-k+1}}&h_{1,u_{n-k+1}}&...&h_{k,u_{n-k+1}}
\\.........&...&...&.........\\g_{u_n}&h_{1,u_n}&...&h_{k,u_n}
\end{array}\right)\,.
\end{equation}
Here
$$\Delta=\det\left(\begin{array}{ccc}h_{1,u_{n-k+1}}&...&h_{k,u_{n-k+1}}
\\.........&...&.........\\h_{1,u_n}&...&h_{k,u_n}
\end{array}\right)$$
and $P_n(g,\xi)$ is given by (\ref{pp}). Notice that $P_{n,k}(h_1,\xi)=...=P_{n,k}(h_k,\xi)=0$.
It is  easy to see that linear transformations of the form $h_i\to
c_{i1}h_1+...+c_{ik}h_k$, $g\to g+d_1h_1+...+d_kh_k$ with constant
coefficients $c_{ij}$, $d_i$ do not change $P_{n,k}(g,\xi)$.

One can verify that
\begin{equation}\label{pp2} (P_{n,k}(g,\xi))_{\xi}=
S_{n,k}(g,\xi)e^{2\pi ir(\tau-\xi)}\frac{\theta^{\prime}(0)^{-s_1-...-s_n}\theta(u_1)^{s_1}...
\theta(u_n)^{s_n}}{\theta(\xi)^{-s_1-...-s_n}\theta(\xi-u_1)^{s_1}...\theta(\xi-u_n)^{s_n}},\end{equation}
where
\begin{equation}\label{ellk}
S_{n,k}(g,\xi)=\frac{1}{\Delta}(\sum_{1\leq \alpha\leq
n-k}\frac{\theta(u_{\alpha})\theta(\xi-u_{\alpha}-\eta)}{\theta(u_{\alpha}+\eta)\theta(\xi-u_{\alpha})}\Delta_{\alpha}(g)-(s_1+...+s_n)\frac{\theta^{\prime}(0)\theta(\xi-\eta)}{\theta(\eta)\theta(\xi)}\Delta_0(g))
\end{equation}
and
$$\Delta_{\alpha}(g)=\det\left(\begin{array}{cccc}g_{u_{\alpha}}&h_{1,u_{\alpha}}&...&h_{k,u_{\alpha}}
\\g_{u_{n-k+1}}&h_{1,u_{n-k+1}}&...&h_{k,u_{n-k+1}}
\\.........&...&...&.........\\g_{u_n}&h_{1,u_n}&...&h_{k,u_n}
\end{array}\right),$$$$\Delta_0(g)=\det\left(\begin{array}{cccc}g&h_1&...&h_k
\\g_{u_{n-k+1}}&h_{1,u_{n-k+1}}&...&h_{k,u_{n-k+1}}
\\.........&...&...&.........\\g_{u_n}&h_{1,u_n}&...&h_{k,u_n}
\end{array}\right).$$

{\bf Proposition 5.} The following relations hold
\begin{equation}
\label{pr3}
(P_{n,k}(g,\xi))_{u_{\alpha}}=-\frac{\Delta_{\alpha}(g)\theta(u_{\alpha})}{\Delta\theta(u_{\alpha}+\eta)}\sum_{n-k+1\leq\beta\leq n}\frac{\theta(u_{\beta}-u_{\alpha}-\eta)(P_{n,k}(g,\xi))_{u_{\beta}}}{\theta(u_{\beta}-u_{\alpha})S_{n,k}(g,u_{\beta})}-
\end{equation}
$$
\frac{\Delta_{\alpha}(g)\theta(u_{\alpha})\theta(\xi-u_{\alpha}-\eta)}{\Delta\theta(u_{\alpha}+\eta)\theta(\xi-u_{\alpha})}e^{2\pi ir(\tau-\xi)}\frac{\theta^{\prime}(0)^{-s_1-...-s_n}\theta(u_1)^{s_1}...\theta(u_n)^{s_n}}{\theta(\xi)^{-s_1-...-s_n}\theta(\xi-u_1)^{s_1}...\theta(\xi-u_n)^{s_n}},$$
where $1\leq\alpha\leq n-k$, and
\begin{equation}
\label{pr4}
(P_{n,k}(g,\xi))_{\tau}=\frac{1}{2\pi i}\sum_{n-k+1\leq\beta\leq n}\frac{(P_{n,k}(g,\xi))_{u_{\beta}}}{S_{n,k}(g,u_{\beta})}(S^{\prime}_{n,k}(g,u_{\beta})-\frac{\theta^{\prime}(-\eta)}{\theta(-\eta)}S_{n,k}(g,u_{\beta}))+\end{equation}
$$
\frac{1}{2\pi i}\Big(S^{\prime}_{n,k}(g,\xi)-
\frac{\theta^{\prime}(-\eta)}{\theta(-\eta)}S_{n,k}(g,\xi)\Big)
e^{2\pi ir(\tau-\xi)}\frac{\theta^{\prime}(0)^{-s_1-...-s_n}\theta(u_1)^{s_1}...\theta(u_n)^{s_n}}{\theta(\xi)^{-s_1-...-s_n}\theta(\xi-u_1)^{s_1}...\theta(\xi-u_n)^{s_n}},$$
where
$$S^{\prime}_{n,k}(g,\xi)=\frac{1}{\Delta}(\sum_{1\leq \alpha\leq
n-k}\frac{\theta(u_{\alpha})\theta^{\prime}(\xi-u_{\alpha}-\eta)}{\theta(u_{\alpha}+\eta)\theta(\xi-u_{\alpha})}\Delta_{\alpha}(g)-(s_1+...+s_n)\frac{\theta^{\prime}(0)\theta^{\prime}(\xi-\eta)}{\theta(\eta)\theta(\xi)}\Delta_0(g)).$$
Moreover, $\displaystyle \frac{(P_{n,k}(g,\xi))_{u_{\beta}}}{S_{n,k}(g,u_{\beta})}$ does not depend on $g$ if $n-k+1\leq\beta\leq n$.

{\bf Proof.} Taking the derivatives of (\ref{pr3}), (\ref{pr4}) with respect to $\xi,$ one arrives at theta-functional identities,
which can be proved straightforwardly. Moreover, the values of the left and the right hand sides of (\ref{pr3}) and (\ref{pr4})
are equal to zero at $\xi=0$.    $\blacksquare$

Let $g_1,~g_2\in {\cal H}$. Assume that $g_1,g_2,h_1,...,h_k$ are
linearly independent. Define pseudopotential
$A_{n,k}(p,u_1,...,u_n,\tau)$ in the parametric form by
\begin{equation}\label{psdef}
A_{n,k}=P_{n,k}(g_1,\xi),\qquad p=P_{n,k}(g_2,\xi).
\end{equation}
To construct  $A_{n,k}(p,u_1,...,u_n,\tau)$, one has to find $\xi$ from the second equation and
substitute into the first one.
The pseudopotential $A_{n,k}(p,u_1,...,u_n,\tau)$ is called {\it elliptic pseudopotential of defect} $k.$

{\bf Theorem 2.} Let $g_0,g_1,...,g_{n-k},h_1,...,h_k\in {\cal H}$
be a basis in ${\cal H}$ and pseudopotentials $B_{\alpha},~\alpha=1,...,n-k$ are
defined by
$$B_{\alpha}=P_{n,k}(g_{\alpha},\xi),\qquad p=P_{n,k}(g_0,\xi), \qquad \alpha=1,...,n-k.$$
Then the compatibility conditions for (\ref{pseudopar}) are
equivalent to the following system of PDEs for $u_1,...,u_n,\tau$:
$$\sum_{1\leq\beta\leq n-k}(\Delta_0(g_q)\Delta_{\beta}(g_r)-\Delta_0(g_r)\Delta_{\beta}(g_q))(u_{\beta,t_s}+\frac{1}{2\pi i}(\frac{\theta^{\prime}(u_{\beta}+\eta)}{\theta(u_{\beta}+\eta)}-\frac{\theta^{\prime}(\eta)}{\theta(\eta)})\tau_{t_s})+$$
\begin{equation}\label{sys3}\sum_{1\leq\beta\leq n-k}(\Delta_0(g_r)\Delta_{\beta}(g_s)-\Delta_0(g_s)\Delta_{\beta}(g_r))(u_{\beta,t_q}+\frac{1}{2\pi i}(\frac{\theta^{\prime}(u_{\beta}+\eta)}{\theta(u_{\beta}+\eta)}-\frac{\theta^{\prime}(\eta)}{\theta(\eta)})\tau_{t_q})+
\end{equation}
$$\sum_{1\leq\beta\leq n-k}(\Delta_0(g_s)\Delta_{\beta}(g_q)-\Delta_0(g_q)\Delta_{\beta}(g_s))(u_{\beta,t_r}+\frac{1}{2\pi i}(\frac{\theta^{\prime}(u_{\beta}+\eta)}{\theta(u_{\beta}+\eta)}-\frac{\theta^{\prime}(\eta)}{\theta(\eta)})\tau_{t_r})=0,$$
$$\sum_{1\leq \beta\leq n-k,\beta\ne
\alpha}\frac{\theta(u_{\beta})\theta(u_{\alpha}-u_{\beta}-\eta)}{\theta(u_{\beta}+\eta)\theta(u_{\alpha}-u_{\beta})}(\Delta_{\alpha}(g_r)\Delta_{\beta}(g_q)-\Delta_{\alpha}(g_q)\Delta_{\beta}(g_r))\times$$$$(u_{\alpha,t_s}-u_{\beta,t_s}+\frac{1}{2\pi i}(\frac{\theta^{\prime}(u_{\alpha}-u_{\beta}-\eta)}{\theta(u_{\alpha}-u_{\beta}-\eta)}-\frac{\theta^{\prime}(-\eta)}{\theta(-\eta)})\tau_{t_s})+$$
$$
\sum_{1\leq \beta\leq n-k,\beta\ne
\alpha}\frac{\theta(u_{\beta})\theta(u_{\alpha}-u_{\beta}-\eta)}{\theta(u_{\beta}+\eta)\theta(u_{\alpha}-u_{\beta})}(\Delta_{\alpha}(g_s)\Delta_{\beta}(g_r)-\Delta_{\alpha}(g_r)\Delta_{\beta}(g_s))\times$$$$(u_{\alpha,t_q}-u_{\beta,t_q}+\frac{1}{2\pi i}(\frac{\theta^{\prime}(u_{\alpha}-u_{\beta}-\eta)}{\theta(u_{\alpha}-u_{\beta}-\eta)}-\frac{\theta^{\prime}(-\eta)}{\theta(-\eta)})\tau_{t_q})+
$$
$$\sum_{1\leq \beta\leq n-k,\beta\ne
\alpha}\frac{\theta(u_{\beta})\theta(u_{\alpha}-u_{\beta}-\eta)}{\theta(u_{\beta}+\eta)\theta(u_{\alpha}-u_{\beta})}(\Delta_{\alpha}(g_q)\Delta_{\beta}(g_s)-\Delta_{\alpha}(g_s)\Delta_{\beta}(g_q))\times$$$$(u_{\alpha,t_r}-u_{\beta,t_r}+\frac{1}{2\pi i}(\frac{\theta^{\prime}(u_{\alpha}-u_{\beta}-\eta)}{\theta(u_{\alpha}-u_{\beta}-\eta)}-\frac{\theta^{\prime}(-\eta)}{\theta(-\eta)})\tau_{t_r})-$$
$$(s_1+...+s_n)\frac{\theta^{\prime}(0)\theta(u_{\alpha}-\eta)}{\theta(\eta)\theta(u_{\alpha})}(\Delta_0(g_q)\Delta_{\alpha}(g_r)-\Delta_0(g_r)\Delta_{\alpha}(g_q))(u_{\alpha,t_s}+\frac{1}{2\pi i}(\frac{\theta^{\prime}(u_{\alpha}-\eta)}{\theta(u_{\alpha}-\eta)}-\frac{\theta^{\prime}(-\eta)}{\theta(-\eta)})\tau_{t_s})-$$
\begin{equation}\label{sys4}(s_1+...+s_n)\frac{\theta^{\prime}(0)\theta(u_{\alpha}-\eta)}{\theta(\eta)\theta(u_{\alpha})}(\Delta_0(g_r)\Delta_{\alpha}(g_s)-\Delta_0(g_s)\Delta_{\alpha}(g_r))(u_{\alpha,t_q}+\frac{1}{2\pi i}(\frac{\theta^{\prime}(u_{\alpha}-\eta)}{\theta(u_{\alpha}-\eta)}-\frac{\theta^{\prime}(-\eta)}{\theta(-\eta)})\tau_{t_q})-
\end{equation}
$$(s_1+...+s_n)\frac{\theta^{\prime}(0)\theta(u_{\alpha}-\eta)}{\theta(\eta)\theta(u_{\alpha})}(\Delta_0(g_s)\Delta_{\alpha}(g_q)-\Delta_0(g_q)\Delta_{\alpha}(g_s))(u_{\alpha,t_r}+\frac{1}{2\pi i}(\frac{\theta^{\prime}(u_{\alpha}-\eta)}{\theta(u_{\alpha}-\eta)}-\frac{\theta^{\prime}(-\eta)}{\theta(-\eta)})\tau_{t_r})=0,$$ where
$\alpha=1,...,n-k$ and
\begin{equation}\label{sys5}
\sum_{\alpha=1}^{n-k}\frac{\Delta_{\alpha}(g_r)\theta(u_{\alpha})\theta(u_{\beta}-u_{\alpha}-\eta)}{\Delta\theta(u_{\alpha}+\eta)\theta(u_{\beta}-u_{\alpha})}u_{\alpha,t_s}-S_{n,k}(g_r,u_{\beta})u_{\beta,t_s}-\frac{1}{2\pi i}(S^{\prime}_{n,k}(g_r,u_{\beta})-\frac{\theta^{\prime}(-\eta)}{\theta(-\eta)}S_{n,k}(g_r,u_{\beta}))\tau_{t_s}=
\end{equation}
$$\sum_{\alpha=1}^{n-k}\frac{\Delta_{\alpha}(g_s)\theta(u_{\alpha})\theta(u_{\beta}-u_{\alpha}-\eta)}
{\Delta\theta(u_{\alpha}+\eta)\theta(u_{\beta}-u_{\alpha})}u_{\alpha,t_r}-S_{n,k}(g_s,u_{\beta})u_{\beta,t_r}-\frac{1}{2\pi i}
(S^{\prime}_{n,k}(g_s,u_{\beta})-\frac{\theta^{\prime}(-\eta)}{\theta(-\eta)}S_{n,k}(g_s,u_{\beta}))\tau_{t_r},$$
where $\beta=n-k+1,...,n$. Here $q, r, s$ run from 0 to $n$ and $t_0=x$.

{\bf Proof} is similar to the proof of Theorem 1.
$\blacksquare$

{\bf Remark 4.} Given $t_{1},t_{2},t_{3},$ Theorem 2 yields a
3-dimensional system of the form (\ref{genern}) with $m=n+1, \, l=n+k+1$
possessing a pseudopotential representation. Indeed,
formula (\ref{sys5}) gives $3k$ linearly independent
equations if $q,r,s=1,2,3$. Formulas (\ref{sys3}), (\ref{sys4}) give $n-k+1$
equations. On the other hand, one can construct exactly $k$ linear
combinations of equations (\ref{sys5}) with
$q,r,s=1,2,3$ such that derivatives of $u_i,~i=n-k+1,...,n$ cancel
out. Moreover, these linear combinations belong to the span of
equations (\ref{sys3}), (\ref{sys4}). Therefore there exist $(n-k+1)+3k-k=n+k+1$
linearly independent equations.

{\bf Remark 5.}  We
have to assume $n\geq k+2$ in (\ref{sys3}), (\ref{sys4}), (\ref{sys5}). Indeed, for $n=k+1$ we cannot construct
more then one pseudopotential and therefore there is no any system
of the form (\ref{genern}) associated with this case. However, the
corresponding pseudopotential generates interesting integrable
(1+1)-dimensional systems of hydrodynamic type (see Section 5).
Probably these pseudopotentials for $k=0,1,...$ are also related to
some infinite integrable chains of the Benney type \cite{fer1,pav1}.

System (\ref{sys3})-(\ref{sys5}) possesses many conservation
laws of the hydrodynamic type. In particular, the following
statement can be verified by a straightforward calculation.

{\bf Proposition 6.} For any $r\ne s=0,1,...,n,$ system (\ref{sys3})-(\ref{sys5}) has $k$
conservation laws of the
form:
\begin{equation}   \label{conlaw}
\left(\frac{\Delta(g_r,h_1,...\hat{i}...h_k)}{\Delta(h_1,...,h_k)}\right)_{t_s}=
\left(\frac{\Delta(g_s,h_1,...\hat{i}...h_k)}{\Delta(h_1,...,h_k)}\right)_{t_r},
\end{equation}
where $i=1,...,k$. Here
$$\Delta(f_1,...,f_k)=\det\left(\begin{array}{ccc}f_{1,u_{n-k+1}}&...&f_{k,u_{n-k+1}}
\\.........&...&.........\\ f_{1,u_n}&...&f_{k,u_n}
\end{array}\right). \qquad \blacksquare$$

Proposition 6 allows us to define functions $z_1,...,z_k$ such that
\begin{equation}   \label{z}
\frac{\Delta(g_r,h_1,...\hat{i}...h_k)}{\Delta(h_1,...,h_k)}=z_{i,t_r}
\end{equation}
for all $i=1,...,k$ and $r=0,1,...,n$.

Suppose $n+1\geq 3k$; then the system of the form (\ref{genern})
obtained from (\ref{sys3})-(\ref{sys5}) with
$q,r,s=1,2,3$ consists of $3k$ equations (\ref{sys5}) (they are equivalent to (\ref{conlaw})) and $n+1-2k$
equations of the form (\ref{sys3}), (\ref{sys4}). Indeed, only $n+1-2k$ equations
(\ref{sys3}), (\ref{sys4}) are linearly independent from (\ref{sys5}). Expressing $\tau,u_1,...,u_{3k-1}$ in terms of
$z_{i,t_1},z_{i,t_2},z_{i,t_3},~i=1,...,k$ from (\ref{z}) and
substituting into $n+1-2k$ equations of the form (\ref{sys3}), (\ref{sys4}), we
obtain a 3-dimensional system of $n+1-2k$ equations for $n+1-2k$
unknowns $z_1,...,z_k,u_{3k},...u_n$. This is a quasi-linear
system of the second order with respect to $z_i$ and of the first
order with respect to $u_j,$ whose coefficients depend on
$z_{i,t_1},z_{i,t_2},z_{i,t_3},~i=1,...,k,$ and $u_{3k},...u_n$. It is clear that the general solution
of the system can be locally parameterized by $n+1-k$ functions in two variables.

In the case $2k\leq n+1<3k$ the functions
$z_{i,t_1},z_{i,t_2},z_{i,t_3},~i=1,...,k$ are functionally
dependent. We have $3k-n-1$ equations of the form
$$R_i(z_{1,t_1},z_{1,t_2},z_{1,t_3},...,z_{k,t_1},z_{k,t_2},z_{k,t_3})=0,\qquad i=1,...,3k-n-1$$
and $n+1-2k$ second order quasi-linear equations. Totally we have
$(3k-n-1)+(n+1-2k)=k$ equations for $k$ unknowns $z_1,...,z_k$. It is
clear that the general solution of this system can be locally
parameterized by $n+1-k$ functions in two variables.

Suppose $n+1<2k$; then  we have  $n+1+k<3k,$ which means that $3k$ equations of
the form (\ref{sys5}) are linearly dependent. Probably in this case the general solution of the system can also be locally
parameterized by $n+1-k$ functions in two variables.

One of the most interesting cases is $n+1=3 k$, when we have a system of $k$ quasi-linear second order
equations for the functions  $z_1,...,z_k$. The simplest case is $k=2.$

\section{Integrable (1+1)-dimensional hydrodynamic-type systems and hydrodynamic reductions}

In this section we present integrable  (1+1)-dimensional
hydrodynamic type systems (\ref{gidra}) constructed in terms of
elliptic hypergeometric functions. These systems appear as the
so-called hydrodynamic reductions of our elliptic pseudopotentials  $A_{n,k}$. Results and formulas of this section
look similar to the rational case (see
\cite{odsok1}).  By integrability of  (\ref{gidra}) we mean the existence of
infinite number of hydrodynamic commuting flows and conservation
laws. It is known \cite{tsar} that this is equivalent to the
following relations for the velocities $v^{i}(r^1,...,r^N)$:
\begin{equation}   \label{semiham}
\partial_{j}\frac{\partial_{i} v^{k}}{v^{i}-v^{k}}=\partial_{i}\frac{\partial_{j}
v^{k}}{v^{j}-v^{k}}, \qquad  i\ne j\ne k. \qquad
\end{equation}
Here $\partial_{\alpha}=\frac{\partial}{\partial r^{i}}, \,
\alpha=1,\dots,N$. The system (\ref{gidra}) is called {\it semi-Hamiltonian}
if conditions (\ref{semiham}) hold.

The main geometrical object related to any semi-Hamiltonian system
(\ref{gidra}) is a diagonal metric $g_{kk},\,k=1,\dots,N$, where
\begin{equation}   \label{metrik}
\frac{1}{2}\partial_{i} \log{g_{kk}} = \frac{\partial_{i}
v^{k}}{v^{i}-v^{k}}, \qquad i\ne k.
\end{equation}
In view of (\ref{semiham}), the overdetermined system (\ref{metrik}) is
compatible and the function $g_{kk}$ is defined up to an arbitrary factor
$\eta_{k}(r^{k})$. The metric $g_{kk}$ is called the {\it metric associated with (\ref{gidra}).}
It is known that two hydrodynamic type systems are compatible
iff they possess a common associated metric \cite{tsar}.

A diagonal metric $g_{kk}$
is called a {\it metric of Egorov type} if for any $i,j$
\begin{equation}   \label{egor}
\partial_i g_{jj}=\partial_j g_{ii}.
\end{equation}
Note that if an Egorov-type metric associated with a
hydrodynamic-type system of the form (\ref{gidra}) exists, then it
is unique. For any Egorov's metric there exists a potential $G$ such
that $g_{ii}=\partial_i G$. Semi-Hamiltonian systems possessing
associated metrics of Egorov type play important role in the theory
of WDVV associativity equations and in the theory of Frobenious
manifolds \cite{dub,kr,pavts}.

 Let
$\tau(r^1,...,r^N),~\xi_1(r^1,...,r^N),...,\xi_N(r^1,...,r^N)$ be a
solution of the system (\ref{gibtsar1}), (\ref{gibtsar11}). It can be easily verified that this
system is in involution and therefore its solution admits a local
parameterization by $2N$ functions of one variable.  Let
$u_1(r^1,...,r^N),...,u_n(r^1,...,r^N)$ be a solution of the
system (\ref{u1}). It is easy to verify that this system is in
involution for each fixed $\beta$ and therefore has an one-parameter family of solutions
for fixed $\xi_{i},~\tau$.

Consider the following system
\begin{equation}\label{gidragen}
r^i_t=\frac{S_{n,k}(g_1,\xi_i)}{S_{n,k}(g_2,\xi_i)}r^i_x,
\end{equation}
where $g_{1},g_{2}$ are linearly independent solutions of
(\ref{elldarbu}), the polynomials $S_{n,k}, \,k>0$ are
defined by (\ref{ellk}), and $S_{n,0}=S_{n}$ (see
(\ref{ell0})).

{\bf Theorem 3.} The system (\ref{gidragen}) is semi-Hamiltonian. The
associated metric is given by
$$g_{ii}=\Big(S_{n,k}(g,\xi)e^{2\pi ir(\tau-\xi)}\frac{\theta^{\prime}(0)^{-s_1-...-s_n}\theta(u_1)^{s_1}...
\theta(u_n)^{s_n}}{\theta(\xi)^{-s_1-...-s_n}\theta(\xi-u_1)^{s_1}...\theta(\xi-u_n)^{s_n}}\Big)^2\partial_i\tau. $$

{\bf Proof.} Substituting the expression for the metric into
(\ref{metrik}), where $v^i$ are specified by (\ref{gidragen}), one
obtains the identity by virtue of (\ref{elldarbu}) and (\ref{gibtsar1})-(\ref{u1}).
$\blacksquare$

{\bf Remark 6.} The system (\ref{gidragen}) does not possess the
associated metric of the Egorov type in general. However, for very
special values of the parameters $s_{i}$ in (\ref{elldarbu}) there exists $g_2\in\cal H$ such that the metric is of
the Egorov type for all solutions of the system (\ref{gibtsar1})-(\ref{u1}).

{\bf Proposition 7.} Suppose that a solution $\xi_1,...,\xi_N,~\tau,~u_1,...,u_n$ of
(\ref{gibtsar1})-(\ref{u1}) is
fixed. Then the hydrodynamic type systems
\begin{equation}\label{gidragencom}
r^i_{t_1}=\frac{S_{n,k}(g_1,\xi_i)}{S_{n,k}(g_3,\xi_i)}r^i_x,~~~~~~~r^i_{t_2}=\frac{S_{n,k}(g_2,\xi_i)}{S_{n,k}(g_3,\xi_i)}r^i_x
\end{equation} are compatible for all $g_1,~g_2$.

{\bf Proof.} Indeed, the metric associated with (\ref{gidragen})
does not depend on $g_2$. Therefore the systems (\ref{gidragencom})
has a common metric depending on $g_3$ and on a solution of
(\ref{gibtsar1})-(\ref{u1}). $\blacksquare$

{\bf Remark 7.} One can also construct some compatible systems of
the form (\ref{gidragencom}) using Proposition 3. Set
$g_2=Z(u_1,...,u_n,u_{n+1},\tau)$ in (\ref{gidragencom}). Here $u_{n+1}$
is an arbitrary solution of (\ref{u1}) (with $n$ replaced by $n+1$) distinct from $u_1,...,u_n$.
It is clear that the flows (\ref{gidragencom}) are compatible for
such $g_2$ and any $g_1\in \cal H$. Moreover, Proposition 3 implies
that the flows (\ref{gidragencom}) are compatible if we set
$g_1=Z(u_1,...,u_n,u_{n+1},\tau)$, $g_2=Z(u_1,...,u_n,u_{n+2},\tau)$ for two
arbitrary solutions $u_{n+1},~u_{n+2}$ of (\ref{u1}).

All members of the hierarchy constructed in Proposition 7 possess a
dispersionless Lax representation of the form
\begin{equation} \label{L}
L _{t}=\{L ,A \},
\end{equation}
where $\{L ,A \}=A_{p}L _{x}-A_{x} L_{p},$
with common
$L=L(p,r^1,...,r^N)$.  Define a function $L(\xi,r^1,...,r^N)$ by the
following system
\begin{equation}\label{L1}
\partial_{\alpha}L=-\frac{1}{2\pi i}\Big(\rho(\xi_{\alpha}-\xi)-\rho(\xi_{\alpha})\Big)L_{\xi}\partial_{\alpha}
\tau,\qquad
\alpha=1,...,N.
\end{equation}
Note that the system (\ref{L1}) is in involution and therefore the
function $L(\xi,r^1,...,r^N)$ is uniquely defined  up to inessential transformations
$L\to\lambda(L)$. To find the function $L(p,r^1,...,r^N)$ one has to
express $\xi$ in terms of $p$  by (\ref{psdef0}) for $k=0$ or by
(\ref{psdef}) for $k>0$.

{\bf Proposition 8.} Let  $u_{1}, \dots, u_{n}$ be arbitrary
solution of (\ref{u1}). Then system (\ref{gidragen}) admits the
dispersionless Lax representation (\ref{L}),
 where $A=A_{n,k}$ is defined by (\ref{psdef0}) for $k=0$ and by
 (\ref{psdef}) for $k>0$.

{\bf Proof.} Define $A=A_{n,k}$ by (\ref{psdef0}) for
$k=0$ and by
 (\ref{psdef}) for $k>0.$ Substituting $A$ into
(\ref{L}) and calculating $L_t$ by virtue of (\ref{gidragen}),
we obtain that (\ref{L}) is equivalent to
$$\partial_iL=\frac{\partial_iP_{n,k}(g_2,\xi)\cdot S_{n,k}(g_1,\xi_i)-\partial_iP_{n,k}(g_1,\xi)\cdot S_{n,k}(g_2,\xi_i)}
{P_{n,k}(g_2,\xi)_{\xi}\cdot
S_{n,k}(g_1,\xi_i)-P_{n,k}(g_1,\xi)_{\xi}\cdot
S_{n,k}(g_2,\xi_i)}L_{\xi}.$$ Taking into account (\ref{pp2}),(\ref{L1})
and writing down $P_{n,k}(g_i,\xi)_{u_1},...,P_{n,k}(g_i,\xi)_{u_{n-k}}$ and $P_{n,k}(g_i,\xi)_{\tau}$ in terms of
$P_{n,k}(g_1,\xi)_{u_{n-k+1}},...,P_{n,k}(g_1,\xi)_{u_n}$ by
(\ref{pr3}), (\ref{pr4}), one can readily verify this equality. $\blacksquare$

Let us show that integrable (1+1)-dimensional systems
(\ref{gidragen}) define hydrodynamic reductions for
pseudopotentials and 3-dimensional systems from Sections 3 and 4.

In \cite{ferhus1, pav2, odpavsok} a definition of
integrability for equations (\ref{L}), (\ref{pseudo}) and
(\ref{genern}) is given in terms of hydrodynamic reductions.

Suppose there exists a pair of compatible semi-Hamiltonian
hydrodynamic-type systems of the form
\begin{equation}\label{dred}
r_{t_1}^{i}=v_1^{i}(r^1,...,r^N)r_{x}^{i}, \qquad r_{t_2}^{i}=v_2^{i}(r^1,...,r^N)r_{x}^{i}
\end{equation}
and functions
$u_i=u_i(r^1,...,r^N)$ such that these functions satisfy (\ref{genern}) for any
solution of (\ref{dred}).  Then
(\ref{dred}) is called {\it a hydrodynamic reduction} for
(\ref{genern}).

{\bf Definition 1} \cite{ferhus1}. A system of the form (\ref{genern}) is called
{\it integrable} if equation (\ref{genern}) possesses
sufficiently many hydrodynamic reductions for each $N\in\N$.
"Sufficiently many" means that the set of hydrodynamic reductions
can be locally parameterized by $2N$ functions of one variable. Note
that due to gauge transformations $r^i\to\lambda_i(r^i)$ we have only $N$
essential functional parameters for hydrodynamic reductions.

Suppose there exists a semi-Hamiltonian
hydrodynamic-type system (\ref{gidra}) and functions
$u_i=u_i(r^1,...,r^N)$, $L=L(p,r^1,...,r^N)$ such that these
functions satisfy dispersionless Lax equation (\ref{L}) for any solution $
r^1(x,t),...,r^N(x,t)$ of the system (\ref{gidra}). Then
(\ref{gidra}) is called {\it a hydrodynamic reduction} for
  (\ref{L}).

{\bf Definition 2} \cite{odpavsok}. A dispersionless Lax equation (\ref{L}) is called
{\it integrable} if equation (\ref{L}) possesses
sufficiently many hydrodynamic reductions for each $N\in\N$.

We also call the corresponding pseudopotential $A(p,u_1,...,u_n)$ integrable.

{\bf Definition 3}  \cite{odsok1}. Two integrable pseudopotentials $A_1,~A_2$ are
called {\it compatible} if the system
$$L _{t_1}=\{L ,A_1 \}, \qquad L _{t_2}=\{L ,A_2\}$$
possesses sufficiently many  hydrodynamic reductions (\ref{dred})
 for each
$N\in\N$.

If $A_1,~A_2$ are compatible, then $A=c_1A_1+c_2A_2$ is
integrable for any constants $c_1$, $c_2$. Indeed, the system
$$r_{t}^{i}=(c_1 v_1^{i}(\mathbf{r})+c_2 v_2^i(\mathbf{r}))\,r_{x}^{i}$$ is a
hydrodynamic reduction of (\ref{L}).

{\bf Definition 4.} By 3-dimensional system associated with compatible
functions $A_1,~A_2$ we mean the system of the
form (\ref{genern}) equivalent to the compatibility conditions for the
system
\begin{equation}\label{sys}
\psi_{t_{2}}=A_{1}(\psi_{t_{1}},u_1,...,u_n), \qquad
\psi_{t_{3}}=A_{2}(\psi_{t_{1}},u_1,...,u_n). \end{equation}

It is clear that any system associated with a pair of compatible
functions possesses sufficiently many hydrodynamic
reductions and therefore it is integrable in the sense of Definition 1.

The following statement is a reformulation of Proposition 8.

{\bf Theorem 4.} The system (\ref{gidragen}) is a hydrodynamic
reduction of the pseudopotential $A_{n,k}$ defined by (\ref{psdef0})
if $k=0$ and by (\ref{psdef}) if $k>0$. Recall that we use the
notation $S_n\equiv S_{n,0},~ A_n\equiv A_{n,0},~ P_n\equiv
P_{n,0}$.

{\bf Proposition 9.} Suppose $g_1,g_2,g_3,h_1,...,h_k\in {\cal H}$
are linearly independent. Define pseudopotentials $A_1$, $A_2$ by
$$A_1=P_{n,k}(g_1,\xi),\qquad A_2=P_{n,k}(g_2,\xi), \qquad p=P_{n,k}(g_3,\xi).$$
Then $A_1$ and $A_2$ are compatible.

{\bf Proof.} Note that the system (\ref{gibtsar1})-(\ref{u1}), (\ref{L1}) does
not depend on $g_1,g_2,g_3$ and therefore we have a family of
functions $L, \xi_i, u_i, \tau$   giving hydrodynamic reductions of the
form (\ref{gidragen}) for both $A_1$ and $A_2$. Moreover, according
to Proposition 7 the systems
$$r^i_{t_1}=\frac{S_{n,k}(g_1,\xi_i)}{S_{n,k}(g_3,\xi_i)}r^i_x, \qquad
r^i_{t_2}=\frac{S_{n,k}(g_2,\xi_i)}{S_{n,k}(g_3,\xi_i)}r^i_x$$ are
compatible. $\blacksquare$

{\bf Remark 8.} This result implies that 3-dimensional hydrodynamic
type systems constructed in Sections 3, 4 possess sufficiently many
hydrodynamic reductions and therefore are integrable in the sence of Definition 1.

{\bf Remark 9.} Using Proposition 3, one can construct compatible
pseudopotentials $A_{1}$ and $A_{2}$ depending on different number of variables $u_i$. Indeed, let
$g_1,g_3,h_1,...,h_k\in {\cal H}$ and $g_2=Z(u_1,...,u_n,u_{n+1},\tau)$.
Then $A_2$ depends on $u_1,...,u_n,u_{n+1},\tau$ and $A_1$ depends on
$u_1,...,u_n,\tau$ only.

\vskip.3cm \noindent {\bf Acknowledgments.} Authors thank B. Feigin, I. Krichever,  M. Pavlov and V. Shramchenko for fruitful discussions.  V.S. was partially supported by the RFBR grants 08-01-464 and NS 3472.2008.2.

\end{document}